# Persistence of superconductivity in niobium ultrathin films grown on R-Plane Sapphire

Cécile Delacour[1,] Luc Ortega[1], Marc Faucher[1,2], Thierry Crozes[1], Thierry Fournier[1], Bernard Pannetier[1] and Vincent Bouchiat[1,*]

[1]*Institut Néel, CNRS-Université Joseph Fourier-Grenoble INP, BP 166, F-38042 Grenoble, France.*

[2]now at *IEMN, CNRS UMR 8520, Avenue Poincaré, BP 60069, 59652 Villeneuve d'Ascq Cedex, France.*

Abstract : **We report on a combined structural and electronic analysis of niobium ultrathin films (from 2 to 10 nm) deposited in ultra-high vacuum on atomically flat R-plane sapphire wafers. A textured polycrystalline morphology is observed for the thinnest films showing that hetero-epitaxy is not achieved under a thickness of 3.3nm , which almost coincides with the first measurement of a superconducting state. The superconducting critical temperature rise takes place on a very narrow thickness range, of the order of a single monolayer (ML). The thinnest superconducting sample (3 nm/9ML) has an offset critical temperature above 4.2K and can be processed by standard nanofabrication techniques to generate air- and time-stable superconducting nanostructures, useful for quantum devices.**

[*]Corresponding author, email : bouchiat@grenoble.cnrs.fr

# Introduction

Superconductivity has been recently shown to survive down to extremely confined nanostructures such as metal monolayers[1] or clusters composed of few organic molecules[2]. While these structures are extremely interesting for probing the ultimate limits of nanoscale superconductivity, their studies are limited to in-situ measurements. Preserving a superconducting state in ultrathin films that can be processed by state-of-the-art nanofabrication techniques and withstand multiple cooling cycles remains technologically challenging and is timely for quantum devices applications. Having the highest critical temperature among elemental superconductors, niobium is an ubiquitous material for superconducting thin films and its performance is known to increase when epitaxy conditions can be reached. Hetero-epitaxy of niobium is best[3, 4] performed on sapphire substrates since both their lattice parameters and thermal dilatation coefficients match rather well[3]. Most combined structural and transport studies have involved A-plane (1000) oriented sapphire[5] as the growing substrate. However the high step-edge roughness[6] and interfacial stress found for that crystal orientation is likely to be a probable source of rapid aging. Only few studies[4, 6] have described Nb growth on R-plane oriented ($1\bar{1}02$) sapphire. However, this orientation provides an ideal substrate to promote epitaxy of Nb with (001) orientation. We present here a combined structural and electronic transport analysis of the early growth stages of niobium films on that orientation which lead to a better understanding of the mechanism of superconductivity depression in ultrathin Nb films and show the existence of an intermediate metallic behaviour.

# Thin-film Growth and Structural Analysis

The niobium films are grown in ultra-high vacuum using electron gun evaporation technique with the following details : R-Plane oriented**,** single faced polished, 2 inches sapphire substrates with reduced miscut (<10$^{-3}$ rad) (from Le Rubis S.A., and Kyocera) are cleaned using water-based surfactants in ultrasonic bath. The surface is then prepared[7, 8] by annealing during 1 hour in air at 1100°C. Wafers are transferred in the deposition chamber (from Isa-Riber, Ultra High Vacuum, base pressure ~10$^{-10}$ torr). Before niobium deposition, in-situ cleaning of the surface is performed during several minutes using Argon milling (Ion tech, Inc) using settings previously reported in ref. [9]. Then, the substrate is heated at 660°C and niobium (ingot target prepared from 99.99% purity pellets from Neyco) is deposited using electron gun at 0.2 to 0.5 nm/s evaporation rate. This substrate temperature has been shown[3] to optimize the deposition conditions as it provides a trade-off between atomic mobility of niobium adatoms on the surface and oxygen contamination migrating from the sapphire that starts to occur above that temperature[5]. The vacuum is kept at residual pressures lower than 10$^{-9}$ Torr during niobium deposition. Finally, the sample is cooled below 80°C and covered by evaporation of a capping layer of 2nm-thick silicon in order to protect the film from oxidation. The thickness $d$ of the niobium layer film is varied from 200 nm down to 2 nm (crystallographic analysis was restricted to the lowest thickness range below 10nm samples A-G).

The sapphire surface annealing that is performed prior to niobium growth has been proven critical to improve the epitaxy[3, 10]. Atomic Force Microscopy analysis performed prior to growth clearly shows vicinal surfaces with atomically flat terraces separated by straight steps edges[7] of 0.3 nm height (Fig. 1a). This measured step height is in agreement with the calculated[4] distance between two adjacent ($1\bar{1}02$) lattice planes ($a_{rhom}$ =0.35 nm) see Fig.1d. After Nb/Si deposition, the surface exhibits similar features replicated on top of the capping layer, which have same height, orientation and size than the bare alumina substrate vicinal surfaces (Fig. 1b). These features along with the low measured roughness indicate a good ordering of the niobium films.

On a larger scale, X-ray scattering measurements confirm an average vicinal surface width averaged over the entire wafer of about 300 nm (Fig. 1c).

In order to probe the inner structure of the multilayer, X-ray reflectivity (XRR) measurements have been performed (Fig. 2a). All samples exhibit a layered structure in agreement with $Al_2O_3$/Nb/Si simulation model (dotted curves) from which a precise determination of the multilayer structure can be deduced (Fig. 2b). The magnitude of the measured roughness at the top and bottom niobium interfaces are deduced from the fits and show almost constant values over the whole deposition range. While the Nb/Sapphire roughness is compatible with an atomically sharp interface, a significantly rougher Nb-Si interface is extracted from the fits (Fig. 2b). However surface roughness measured by AFM (Fig. 1b) on the top of the silicon capping layer is repeatedly found below 0.1 nm. This suggests the presence of a diffuse and smooth Nb/Si interface rather than a sharp one, which might result from a slight inter-diffusion zone of about 2 monolayers (ML). X-ray diffraction (XRD) measurements reveal that films thicker than 4 nm (samples A to C) are single crystalline with a (001) niobium bulk orientation in agreement with the models (Fig. 1d). The (110) pole figure (Fig. 2c) shows the diffracted intensities of the 4 diagonal (110)-planes of the body center cubic lattice of niobium (Eulerian angle $\chi = 45°$ +/- 3°) as expected for a hetero-epitaxial growth of (001) on R-plane sapphire (Fig. 1c). The 3° angle mismatch corresponds to a previously reported[4] misalignment between the ($1\bar{1}02$) sapphire and (001) niobium planes. Interestingly, this mismatch is observed even for the thinnest single-crystalline sample (4 nm/ equivalent to 12 ML). XRD patterns of samples thinner than 4nm (samples D, E, F) exhibit no trace of the body center cubic Nb lattice, but reveal only a hexagonal lattice corresponding to the sapphire substrate, suggesting that hetero-epitaxy is not achieved if the growth is interrupted in its early stages.

Further investigations by grazing incidence X-ray diffraction confirm that the thinnest films

(samples D, E) are indeed made of mosaic polycrystals, as the patterns are similar to the ones encountered for niobium powder (Fig. 2d). The presence of polycrystals most probably explains the slight discrepancy observed for two samples between XRR measurements and the fit (Fig. 2a). Indeed the model involves three layers of single crystals.

All XRR measurements agree with a structural transition from polycrystalline for the thinnest films to a single crystalline growth occurring for films thicker than 3.3 nm (10 ML).

The initial growth stage of niobium films on an atomically flat substrate follows the step-flow model[11] as previously predicted[4]. Niobium adatoms move on the terraces, rearrange themselves along the sapphire step-edges and form patches which eventually percolate, leading to a mosaic polycristal film. When the layer gets thicker, typically over 10 MLs, volume interactions become dominant with respect to step-edge interaction and force the film to be single-crystalline with a (100) niobium bulk orientation.

## Electronic Transport Measurements

The sheet resistance of samples have been measured from room temperature down to 0.1 K on 5 micron-wide strip-lines over a length of typically 100 squares. Sheet resistance measured as a function of the temperature (Fig. 3a) shows superconducting transitions for all films thicker than 3 nm.

As expected, the critical temperature $T_c$ (defined by default as the temperature at which the film has lost half of its resistance) is found to decrease simultaneously to the film thickness $d$. The superconducting transition widens notably for the polycrystalline films (samples D, E). This transition widening mostly occurs at the top of the transition as a result of an increase of the superconducting fluctuations. However there is still a rather sharp drop of several orders of magnitude of the film resistance at $T_c$ (fig 3a, right) and we found no trace of residual resistance below $T_c$ that would have revealed the existence of phase fluctuations. As for the thinnest films (samples F,G), one does not observe any superconducting state at the lowest temperatures (0.1K) reached in our filtered measurement setup. This suggests that the superconductivity is suppressed for a critical thickness $d_c$ ranging between 2.8 and 3 nanometers.

For sample F however, high resolution measurements have shown a drop at 0.5K towards a partially resistive state which equals 70% of the resistance measured at 5K, (see Fig. 3c). This suggests that the sample experiences a transition towards an intermediate metallic state.

Interestingly, when one crosses the critical thickness $d_c$, one can simultaneously observe a change of sign of the temperature coefficient of resistance *(dR/dT)* measured below 70K as shown in figure 4. In samples F and G which lie below $d_c$, the conductance decreases upon cooling. For sample G, this decrease clearly follows a logarithmic dependence with temperature (Fig.4) , which is a signature of a 2D weak localization.

It is not clear, however, for sample G, whether this transition is towards a metallic state or an insulating one, as both systems are known to exhibit negative temperature coefficient[12].

The magnetic field dependence of the R(T) curve has been measured as well for the 3nm film (Fig. 3b, sample D) and similarly shows a rapid drop of the superconducting transition for fields above 3T together with a slight increase of the resistance upon cooling. The same curve measured for sample F (Fig. 3c) shows a very rapid suppression of the partial drop at 0.5K under a small transverse magnetic fields and a similar change of sign of the temperature coefficient for higher fields.

The purity of Nb films and its thickness-dependence can be estimated by measuring the residual resistivity ratio (RRR) which is obtained by dividing the room temperature resistance by the resistance $R_n$ taken just above the transition. It is known that the quantity *k=(RRR-1)^{-1}* is proportional to the density of defects[13] . The RRR extracted from R(T) curves show a linear dependence on the thickness $d$ (Fig. 5a) and the quantity $k$ diverges for the thinnest samples (Fig. 5b), revealing a high concentration of defects within the top Nb monolayers. One can also deduce from the resistance-temperature measurement the mean free path $l$ knowing that its product with the bulk resistivity ($\rho$) is equal[14] to $\rho l = 3.75 \times 10^{-6}\ \mu\Omega.cm^2$. One finds $l = 3.5$ nm for the 5 nm film. One can then evaluate the superconducting coherence length in the dirty limit[14] : $\xi = 0.852\sqrt{\xi_0 \ell}$ , with Nb-bulk coherence length $\xi_0$=38 nm. One obtains $\xi$=9.8 nm for the 5-nm-thick sample. The superconducting coherence length equals twice the film thickness $d$ and clearly places our study within the two dimensional limit.

A more accurate determination of the superconducting coherence length ($\xi$) can be extracted[15] by measuring the temperature dependence of the critical magnetic field $H_{c2}$(T) (Fig.6) at small applied fields.

For that purpose, we have measured $R$(T) curves for increasing magnetic fields. A precise determination of the transition temperature $T_c$ is obtained by fitting the initial drop of resistance above $T_c$ with a model describing the superconducting fluctuations defined by the quantity $(R^{-1} - R_n^{-1})^{-1}$, measured in presence of a weak magnetic field. Superconducting fluctuations have been calculated by Ullah and Dorsey[16] in the limit of 2D thin films[17]. Above the critical temperature, the fluctuation term follows a linear function of $T - T_c(H)$ which allows to calculate $T_c(H)$ by extrapolating the fitting line. By inverting the set of $T_c$ obtained by these fits, one can then plot the $H_{c2}$(T) curve in the low field limit (inset of Fig.6). Since the Ginzburg-Landau theory gives : $H_{c2}(T)=\frac{\Phi_0}{2\pi\xi^2}(1 - \frac{T}{T_c})$, a linear fit of this curve leads to a determination of the coherence length $\xi$=10.6 nm, which is in good agreement with the first estimate.

## Physical origins of the depression of the superconductivity

The suppression of superconductivity has been studied before in several configurations, for granular[18] or amorphous films[19] and with different elemental materials[19] and alloys[20]. Its origins could be multiple as they range from transition to a localized state due to disorder[21-23] to quantum confinement[18, 24] or to inverse proximity effect with the surrounding materials[25, 26]. In a disorder induced superconducting to insulator transition[27], $R_s$ was believed to be of the order of the resistance quantum $h/(4e^2)$=6.5 kΩ. In our system however the transition is observed for a well lower resistance per square (Fig. 3a) suggesting that the localization is not the dominant factor that leads to the suppression of the superconducting state.

## Finkelstein's model

We analyze our results in the framework of a weak disorder model in homogeneous superconducting films established by Finkel'stein[21]. The suppression of superconductivity is driven by impurities that reinforce Coulomb and spin interactions. $T_c$ is expressed as a function of sheet resistance $R_S$ and the elastic diffusion time $\tau$ :

$$\frac{T_c}{T_{c0}} = e^{\gamma}\left(\frac{1/\gamma - \sqrt{t/2} + t/4}{1/\gamma + \sqrt{t/2} + t/4}\right)^{1/\sqrt{2t}} \tag{1}$$

with $t = \frac{R_S e^2}{\pi h}$, $\gamma = \ln\left(\frac{h}{k_B T_{c0} \tau}\right)$ and $T_{c0}$ = 9.22 K is the Nb-bulk critical temperature. The transition occurs for a non-universal sheet resistance such as $t=\gamma^2$. Fig. 7 compares the Finkel'stein model (Eq. 1) with our experimental data for several values of the γ parameter. The model seems at first sight rather consistent with our data measured above $d_c$ and leads to a fitting parameter γ= 9.5. However, the model is in complete disagreement with our data obtained below the critical thickness $d_c$, as it would imply a much smoother superconducting depression. Furthermore the model implies a critical sheet resistance $R_c$ = 900 Ω at the transition, which is twice the experimental one measured at $d_c$ (Fig. 3a). Furthermore, in order to get a critical resistance of 500 Ω, the elastic relaxation time τ one has to inject in Eq. (1) would have to be less than $1.5.10^{-17}$ s which gives $v_F\tau$=0.17 nm, a value in disagreement with the mean free path evaluated above.

**Inverse proximity effect induced by the capping layer.**

The formation of metallic alloys and the interdiffusion of silicon or oxygen atoms at the interface is likely to be involved in the disappearance of the superconductivity by an inverse proximity effect for the thinnest films, and their influence is probably aggravated by the fact that the film cannot be grown as a single crystal for thicknesses below 4 nm. According to the simplified McMillan model [28], we consider an inverse proximity effect in a planar thin film geometry. The critical temperature $T_c$ is depressed as a function of the normal layer thickness $d_N$ according to the relation :

$$T_c = T_{c0} \left( \frac{3.56 \cdot T_D}{T_{c0} \cdot \pi} \right)^{-\frac{\alpha}{d}} \qquad (2)$$

with $\alpha = d_N \cdot N_N(0)/N_S(0)$

$N_{N,S}(0)$ are respectively the density of states in normal (N) and superconducting (S) layers and $T_D$ =277 K is the Nb Debye temperature. We have compared our measurements with (Eq. 2) as a function of α (Fig. 8) and find the best agreement to be obtained for $d_N$ =0.55 +/-0.05 nm, considering a $N_N(0)/N_S(0)$ ratio equal to 1. This value is consistent with the thickness of the Nb/Si interdiffusion layer estimated from the XRD data (Table in Fig. 2b Si-interface thickness), which could create a metallic silicide at the Nb/Si interface. Oxygen traces migrating through the silicon capping layer could also further contribute to the inverse proximity effect since niobium oxides with low stoichiometry are known to be metallic and even suspected to be a potential source of destruction of superconductivity by magnetic effects[29].

In addition, when compared with previous studies[30], this simple model is well consistent with our data above $d_c$. However the drastic fall of $T_c$ for thinnest resistive samples (red stars) is not explained even by taking more realistic models[31, 32]. The reason why these models cannot describe the full set of data is likely to be associated with the structural transition in the film, as the physical origin of the superconducting depression most probably varies according to the crystalline structure of the films. Additional models are therefore needed to describe the situation for the thinnest films.

**Proximity effect between 2D grains within the layer.**

One has shown in the growth section that the film morphology differs considerably below the critical thickness $d_c$. Mesoscopic fluctuations of the metal properties can create huge inhomogeneities leading to normal and superconducting regions close to each other. The appropriate model to describe the electronic structure is therefore not anymore the McMillan model but rather models involving Normal/Superconducting 2D networks. Such a model implies a much more drastic effect on the superconducting depression than the one found using the McMillan model.

In the case of a transparent N/S interface one finds the Cooper limit $T \propto e^{-\frac{d_N}{d}}$. Spin flip interactions would even have a more detrimental effect on $T_c$. This latter model would then involve a transition from superconductivity towards a "bad" metal which square resistance indeed does not require a square resistance close to $h/4e^2$ to be triggered.

Below 4 nm, films have been found to be polycrystalline (Fig 2d). Macroscopic superconducting state can be ensured by proximity effect between Nb-grains. Such 2D NS granular network systems have been investigated by Feigel'man et al.[33] but this model concerns the limit of diluted superconducting grains, which is unlikely to apply here. The approach of Spivak et al.[34] needs to be investigated in further details as it deals with a situation closer to our system. It indeed focuses on superconducting grain size (R<ξ) and film thickness (d<ξ) lower than the superconducting coherence length. In both models, it is worth noting that the critical resistance at the transition may appear at arbitrarily small values as it is experimentally observed here.

**Applications to the realization of Quantum Devices**

The reported single crystalline niobium films are stable in time over months even exposed to the ambient atmosphere, as demonstrated by the 4-nm-thick film having a critical temperature $T_c$= 5 K that is still preserved after multiples thermal cycles and 3 years of shelf storage. The thinnest samples are also resistant to micro and nano-fabrication processes, with a superconducting transition temperature preserved above 4.2K for nanostructures made from sample C.

Elaborating time- and air-stable ultrathin superconducting films is useful for quantum device fabrication. An ultrathin film improves the sensitivity of single particle detectors (either based on bolometric, kinetic inductance, or hot spot diffusion effects). These quantum detectors are in high demand for astrophysics particle detectors or for quantum cryptography applications[35]. These films have already proven to be useful for fabricating more exploratory devices such as single electron transistors that have shown room temperature operation[36] or nanoscale magnetometers[37] allowing for increased detection sensitivity of nanomagnet magnetization[38]. Growing high quality niobium (Nb) films has also shown to increase the stability and supercurrent density of Nb-based tunnel junctions[39]. When applied to superconducting single-photon detection, ultrathin Nb-based devices have provided faster repetition times[40] than niobium nitride counterparts due to their lower kinetic inductance. Finally the resilience of such films upon the application of in-plane magnetic field offers promising perspectives of applications including the realization of on-chip superconducting LC resonators that remain superconducting upon intense magnetic fields. This would help for the manipulation of spin qubits coupled to readout superconducting cavities (for a review see ref. [41]).

In order to demonstrate that it is possible to produce stable nanostructures following our technique, the 3-nm-thick (9 ML) Nb-thick superconducting film has been patterned using conventional Deep-UV photolithography followed by reactive ion etching (using $SF_6$). Critical temperature measured on micron wide lines is barely affected by the process and local anodisation induced by Atomic Force microscope based nanolithography lead to the thinnest nanometer scaled integrated on-chip nanodevices ever made so-far[36,37].

**Conclusion**

We have presented a combined analysis of structural and electron transport properties of sub-10 nm niobium films evaporated in ultra-high vacuum on R-plane sapphire. Nb films undergo a sharp drop of the superconducting transition at 3 nm (9 ML) accompanying a structural transition from poly to single crystal occurring at a thickness slightly above that value. The *Tc* cut-off occurs over a narrow thickness range (1ML) obtained for a non-universal critical resistance which is more than 10 times below the quantum of resistance *$h/(4e^2)$*. Resistance temperature measurements show a transition toward a metallic state with 2D weak localization features. Above that critical thickness, the overall stability demonstrated by both structural and electronic measurements of the ultrathin films offers a reliable starting material for the realization of quantum superconducting devices.

**Acknowledgements.**

The authors are deeply indebted to G. Battuz and A. Hadj- Azzem for substrate preparation. One of us (C.D.) thanks the D.G.A. for grant support.

*Correspondence and requests for materials should be addressed to V.B.(vincent.bouchiat@grenoble.cnrs.fr)*

# Figure captions

**Figure 1**

**a**, Atomic force micrograph of an annealed R-plane sapphire (see methods), showing 300 nm wide and 0.3 nm high atomic steps and vicinal surfaces (indicated with the arrow). Scale bar is 0.5 µm.

**b**, Atomic force micrograph of a 5-nm-thick-Nb stripline. The Nb film has been deposited on the substrate shown in 1.a and the stripline is seal realized by photolithography and etching. Arrows indicate the position of the substrate step edges replicated on top of the niobium stripline. Scale bar is 0.5 µm.

**c**, X-ray off-specular scattered intensity versus the incident angle performed on a pristine niobium/silicon film deposited on R-plane sapphire. Positions of the two satellite peaks (of lower intensity) result from diffraction effects induced by atomic steps. The extracted terrace width (300nm, L=mλ/(cosΘ$_1$-cosΘ$_2$)) is in agreement with the one extracted from the AFM micrographs. (X-ray spot size is of the order of 1 mm²).

**d,** Schematics showing the matching of the boby center cubic (bcc)-Nb lattice and the hexagonal-sapphire lattice most commonly represented in the rhombohedra lattice. The (001)-Nb plane follows epitaxy onto ($1\bar{1}02$)$_{rhom}$-sapphire plane (a$_{Nb}$=0.33 nm; a$_{rhom\text{-}sapphire}$ =0.35 nm).

**Figure 2 `**

**a**, X-ray reflectivity (XRR) performed on ultrathin Nb films of decreasing thickness (A: 27 ML, B: 24 ML, C: 12 ML, D: 10 ML, E: 9 ML). The fit (dashed lines) involves a three-layer stack model which involves the capping layer (Al$_2$O$_3$–Nb–Si ).

**b**, Table showing the physical parameters of the tested samples labelled from A to G. The Nb-Si thicknesses and top/bottom-Nb interfacial roughness (Si/Al$_2$O$_3$-Nb) are deduced from the three-layer stack model shown in a. This measurement is not shown for the thinnest sample (sample G) as it lies below the detection threshold of the instrument. Bottom rows underline the two critical thicknesses characterizing both the structural and electronic transitions respectively occurring at 12 ML and at around 9 ML.

**c,** XRD (110)$_{Nb}$-pole figure performed sample C (4 nm film), showing the diffraction of the 4 diagonal (110)$_{Nb}$-plans ($\chi=45°$) of the Nb body centered cubic structure. The (100)$_{Nb}$-plane is shown to be parallel with the Sapphire R-plane in agreement with a hetero-epitaxial growth. Data were corrected for background signal.

**d,** Grazing incidence X-ray diffraction analysis performed on a series of films with different Nb thicknesses (reported in Fig2 b). Curves are shifted for sake of clarity. To illustrate the difference between single and polycrystalline films, the bottom inset show the diffraction pattern of a thicker polycrystalline niobium film (A': 10nm) (red curve) and compares it to a theoretical pattern of niobium powder (black curve).

**Figure 3**

**a** Sheet resistance $R_s$ of niobium films measured as a function of the temperature (at left, temperature in log scale, at right, resistance in log scale). As the thickness $d$ is reduced (from A to G), the residual resistance at 10 K increases and the superconducting transition temperature ($T_c$) is depressed compared to the bulk value ($T_{co}$, indicated with arrow). Note the very narrow thickness range on which the superconductivity depression occurs.

**b** Sheet resistance of sample D (3.3 nm), measured for increased transversal magnetic fields (from 0 to 3 T, with steps of 0.5T).

**c** Normalized resistance of sample F ($d = 2.8$ nm), measured at very low temperature. From bottom to top, the applied transversal magnetic field is respectively 0, 0.35T, 6.5 T.

**Figure 4**

Temperature dependence of the sheet conductance of the thinnest samples (E: d=2.98 nm; F: d=2.8nm; G: d=2 nm) shown in a semilog plot. The conductance of sample G shows a logarithmic decrease upon cooling while for sample F a partial transition occurs at $T \sim 0.5K$.

**Figure 5 a.** Residual resistivity ratio (RRR) as a function of Nb films thickness (black squares) underlined by a linear fit (black curve). RRRs of 200 and 100 nm Nb-thick samples are added in inset to show the asymptotic behaviour towards the bulk regime. **b.** Plot of $k=(RRR-1)^{-1}$ showing the divergence observed at the thinnest thickness.

**Figure 6**

Critical magnetic field as a function of temperature for a 5nm thick Nb film (hollow squares) and for a 3 nm thick Nb film (black squares) obtained by determining the critical temperature from R(T,H) curves. Solid lines are just guides for the eyes. Near Tc(H=0), the critical temperature is obtained by a fit using the Ullah-Dorsey theory (see text).
 Inset: zoom of Hc(T) near $T_c$ for the 5 nm thick film. The linear fit (black curve) gives for the coherence length ξ=10.6 nm.

**Figure 7**

Plot of the ratio, $T_c/T_{co}$ as a function of the film sheet resistance $R_S$ for the whole set of ultrathin films. $R_S$ is measured above the superconducting transition (10K). Data (black squares) are compared with the model of Finkel'stein in weak disorder homogeneous superconducting films for γ=9 and 10 (black curves). Sheet resistances of films below the critical thickness are also reported with an effective $T_c$ arbitrarily taken at the lowest temperature achieved by the experiment (red stars).

**Figure 8**

Critical temperature $T_c$ measured as a function of 1/d for several Nb-thicknesses: from 200nm to 2nm (black squares). Data are compared with the Mc Millan model for several values of the parameter α (black curves). Sheet resistances of thinnest films that do not show superconductivity are also reported (red stars).

REFERENCES

[1] T. Zhang, et al., Nature Physics **6**, 104 (2010).
[2] K. Clark, A. Hassanien, S. Khan, K.-F. Braun, H. Tanaka, and S.-W. Hla, Nature Nanotech **5**, 261 (2010).
[3] G. Oya, M. Koishi, and Y. Sawada, Journal of Applied Physics **60**, 1440 (1986).
[4] A. Wildes, J. Mayer, and K. Theis-Bröhl, Thin Solid Films **401**, 7 (2001).
[5] C. Sürgers and H. Löhneysen, Appl. Phys. A **54**, 350 (1992).
[6] B. Wolfing, K. Theis-Brohl, C. Sutter, and H. Zabel, J. Phys. Condens. Matter **11**, 2669 (1999).
[7] T. Nguyen, D. Bonamy, L. Van, L. Barbier, and J. Cousty, Surface Science **602**, 3232 (2008).
[8] Z. Wang and J. Bentley, Ultramicroscopy **51**, 64 (1993).
[9] S. I. Park, A. Marshall, R. H. Hammond, T. H. Geballe, and J. Talvacchio, Journal of Materials Research, MRS **2**, 446 (1987).
[10] Y. Gotoh, K. Matsumoto, and T. Maeda, Jap. J. Appl. Phys. **41**, 2578 (2002).
[11] C. Flynn, J. Phys. F: Met. Phys. **18** L195 (1988).
[12] S. Maekawa and H. Fukuyama, J. Phys. Soc. Jpn. **51**, 1380 (1982).
[13] C. Kittel, Introduction to Solid State Physics John Wiley & Sons (1976).
[14] N. W. Ashcroft and N. D. Mermin, Solid State Physics, Ed. Holt, Rinehart, and Winston (1976).
[15] M. Tinkham, Introduction to superconductivity, Dover, N-Y. (2004).
[16] S. Ullah and A. T. Dorsey, Phys. Rev. B **44**, 262 (1991).
[17] M. H. Theunissen and P. H. Kes, Phys. Rev. B **55**, 15 183 (1997).
[18] S. Bose, R. Banerjee, and A. Genc, J. Phys. Condens. Matter **18**, 4553 (2006).
[19] N. Nishida, S. Okuma, and A. Asamitsu, Physica B: Condensed Matter **169** 487 (1991).
[20] L. Dumoulin, L. Bergé, J. Lesueur, H. Bernas, and Chapellier M., J. of Low temp. Phys. **93**, 301 (1993).
[21] A. Finkel'Stein, Physica B: Condensed Matter **197**, 636 (1994).
[22] B. Sacépé, C. Chapelier, T. Baturina, and V. Vinokur, Physical Review Letters **101**, 157006.1 (2008).
[23] Y. Dubi, Y. Meir, and Y. Avishai, Nature **449**, 876 (2007).
[24] Y. Guo, et al., Science **306**, 1915 (2004).
[25] O. Bourgeois, A. Frydman, and R. C. Dynes, Phys. Rev. B **68**, 092509.1 (2003).
[26] J. W. P. Hsu, S. I. Park, G. Deutscher, and A. Kapitulnik, Phys. Rev. B **43**, 2648 (1990).
[27] A. M. Goldman and N. Markovic, Physics Today **51**, 39 (1998).
[28] W. L. McMillan, Phys. Rev. **175**, 537 (1968).
[29] T. Proslier, J. F. Zasadzinski, L. Cooley, C. Antoine, J. Moore, J. Norem, M. Pellin, and K. E. Gray, Appl. Phys. Lett. **92**, 212505 (2008).
[30] S. I. Park and T. H. Geballe, Physica B: Condensed Matter **135**, 108 (1985).
[31] W. Silvert, Phys. Rev. B **12**, 4870 (1975).
[32] W. L. McMillan, Phys. Rev. **167**, 331 (1968).
[33] M.V.Feigel'man, A.I.Larkin, and M.A.Skvortsov, Phys. Rev. Lett. **86**, 1869 (2001).
[34] B. Spivak, A. Zyuzin, and M. Hruska, Phys. Rev. B **64**, 132502 (2001).
[35] C.M.Natarajan, et al., Quantum Commmunication and Quantum Networking,Lectures Notes of Institute for Computer Sciences, Springer **36**, 225 (2010).
[36] J. I. Shirakashi, K. Matsumoto, N. Miura, and M. Konagai, Appl. Phys. Lett. **72**, 1893 (1998).

37. Bouchiat V., Faucher M., Thirion C., Wernsdorfer W., Fournier T., and Pannetier B., Appl. Phys. Lett. **79**, 123 (2001).
38. M. Faucher, P. O. Jubert, O. Fruchart, W.Wernsdorfer, and V. Bouchiat, supercon. Sci. Tech. **22**, 064010 (2009).
39. S. Celaschi, T. H. Geballe, and W. P. Lowe, Appl. Phys. Lett. **43**, 794 (1983).
40. A. J. Annunziata, D. F. Santavicca, J. D. Chudow, L. Frunzio, M. J. Rooks, A. Frydman, and D. E. Prober, IEEE Trans. Appl. Superconductivity **19**, 327 (2009).
41. T. Duty, Physics **3**, 80 (2010).

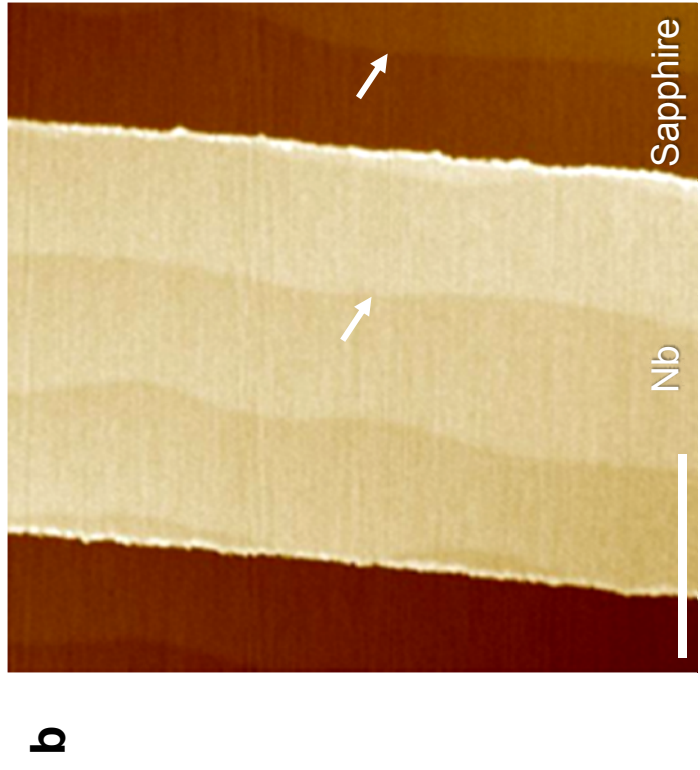 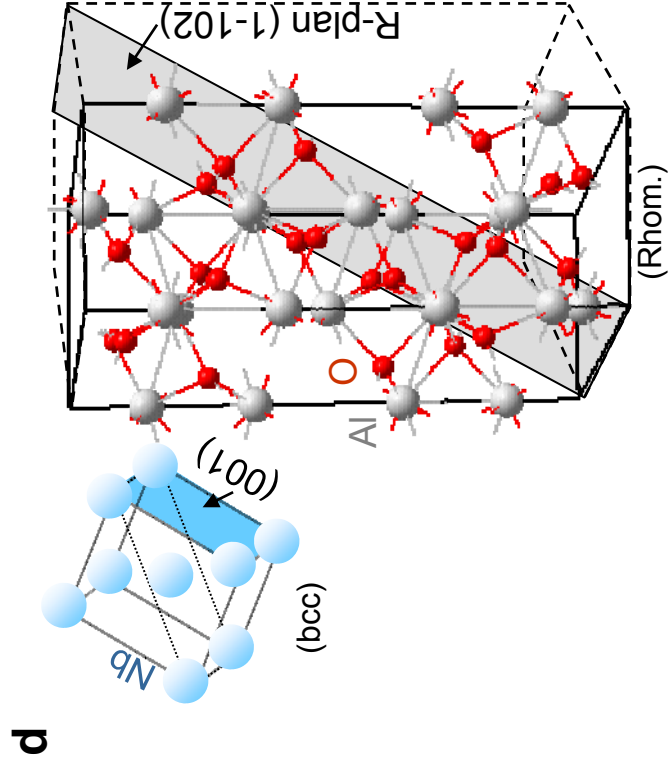

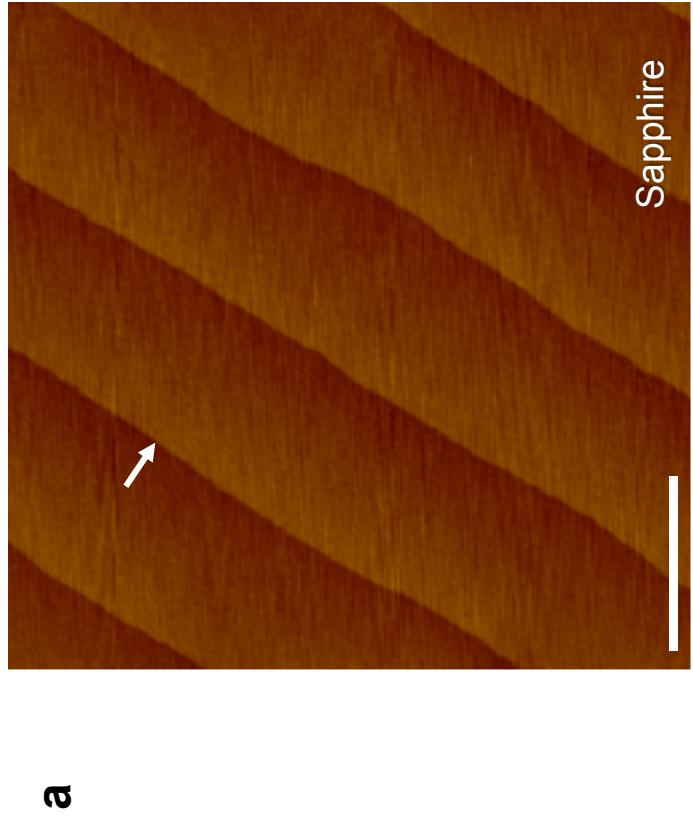 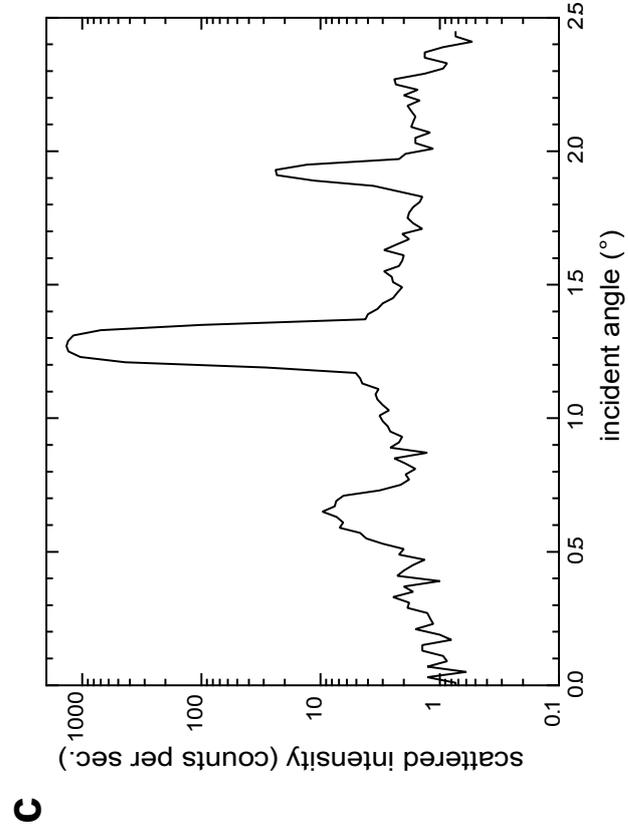

Figure 1    Delacour et al.

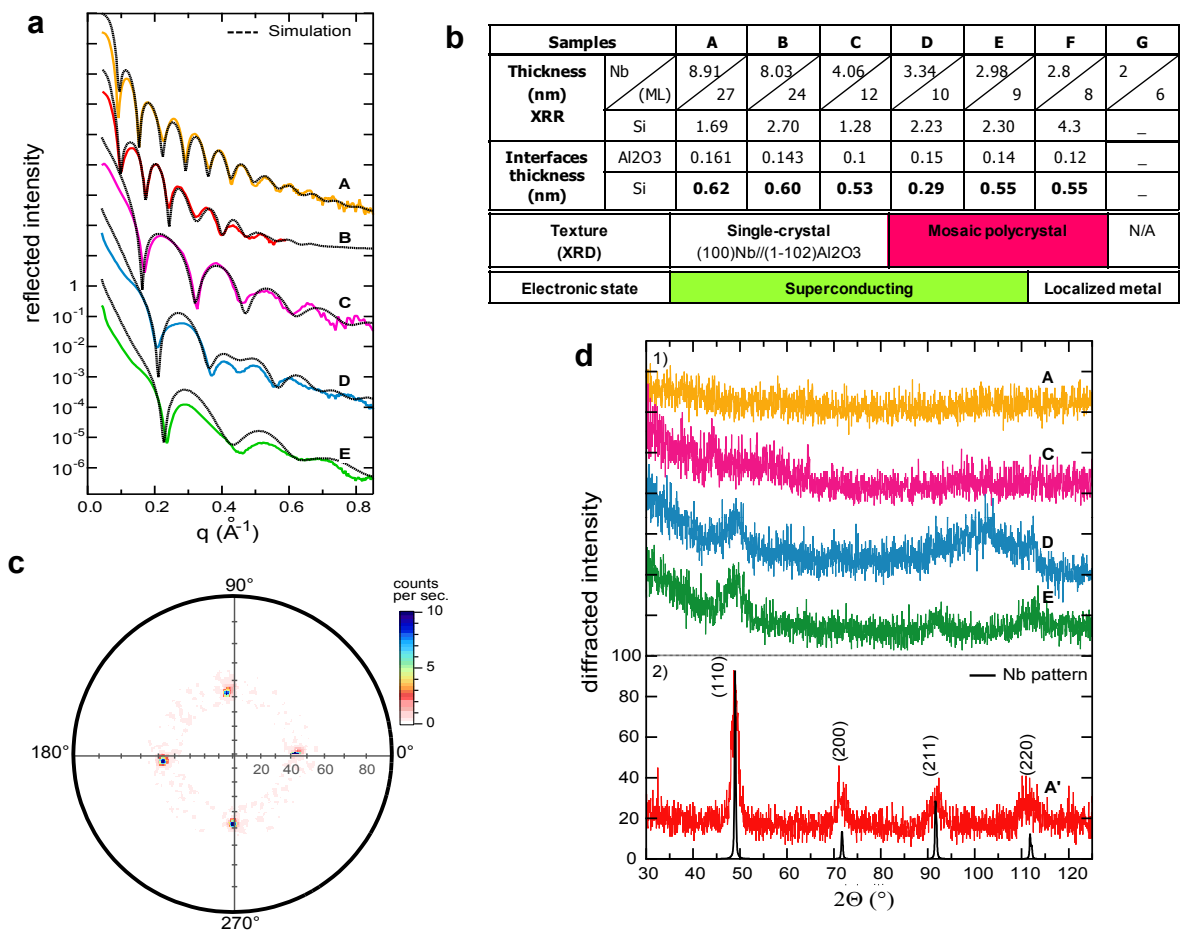

Delacour et al. Figure 2

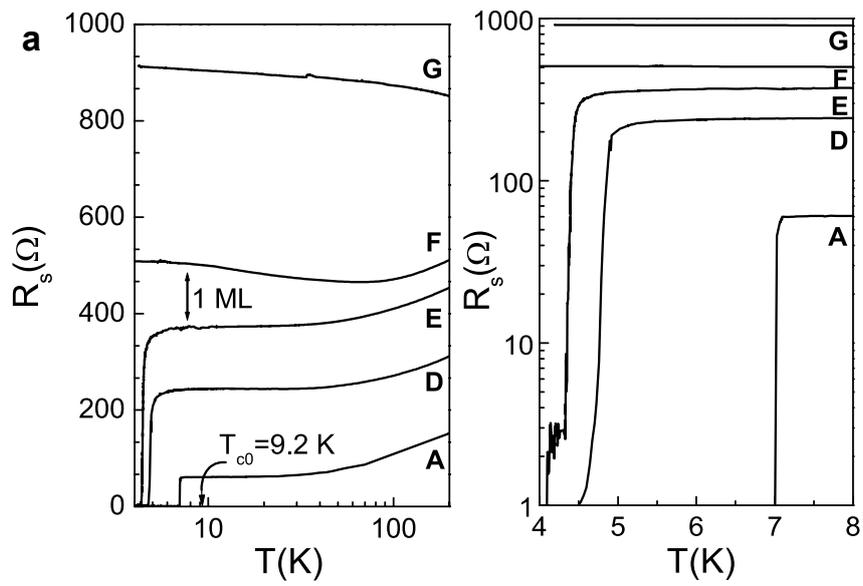

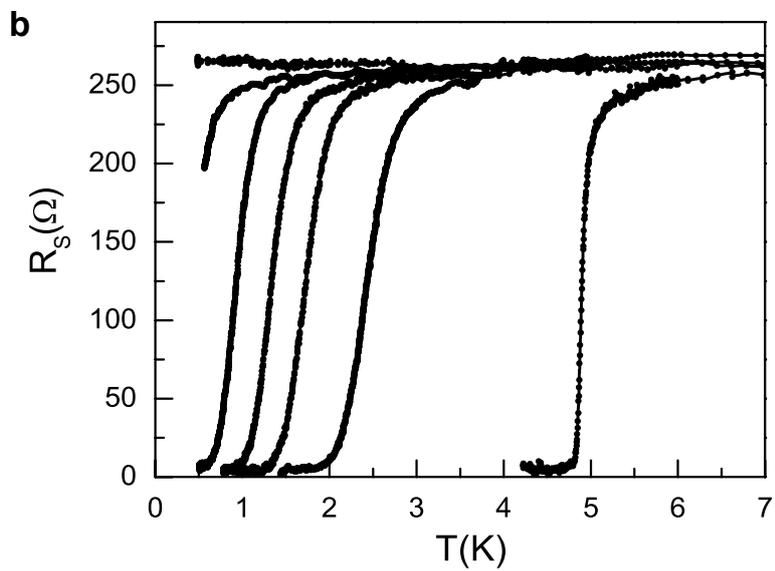

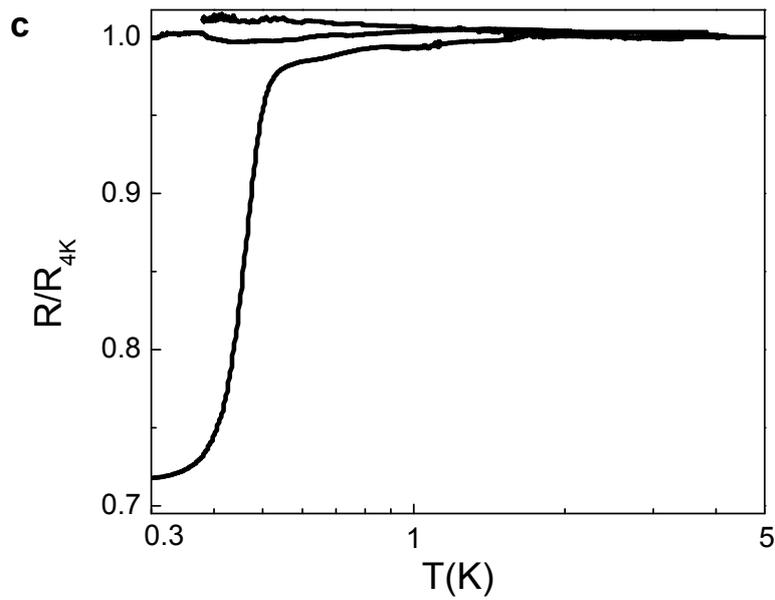

Delacour et al. Figure 3

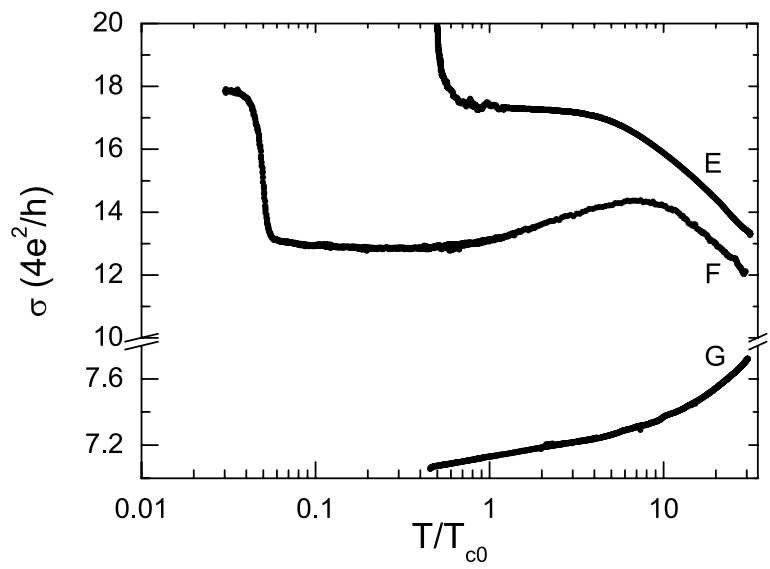

Delacour et al. Figure 4

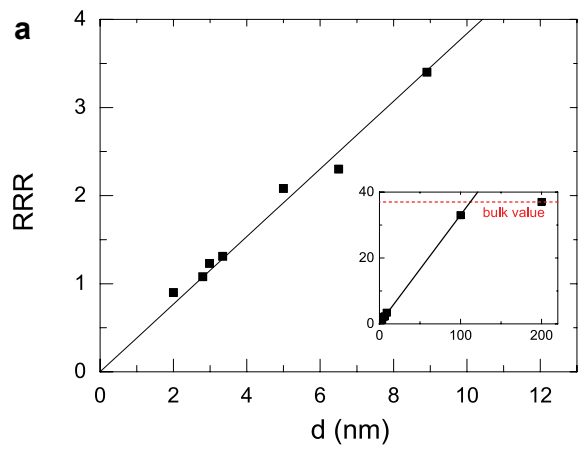
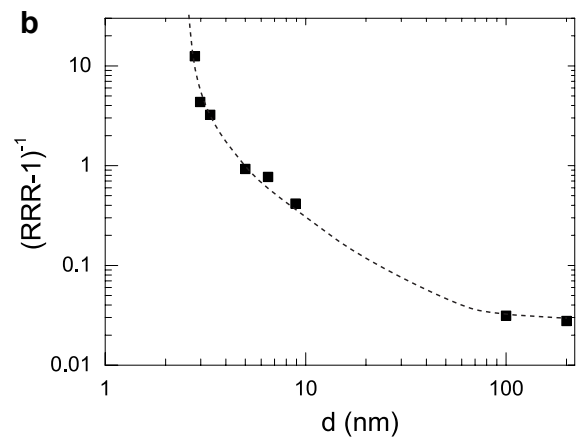

Delacour et al. Figure 5

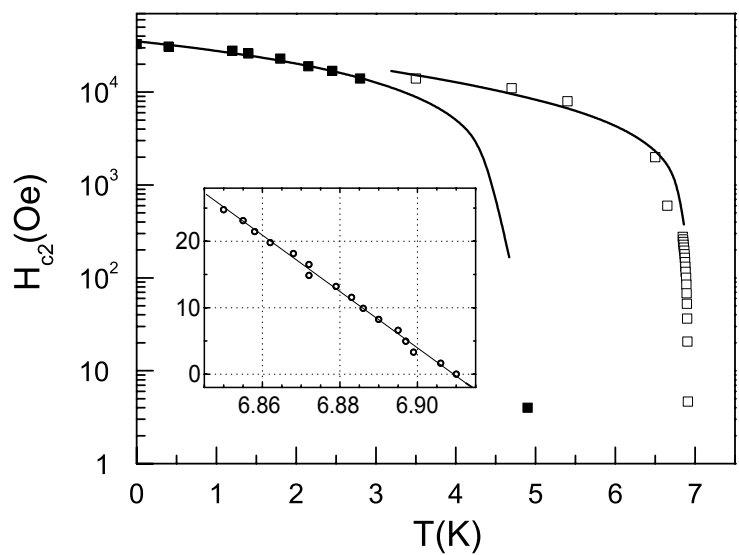

Delacour et al. Figure 6

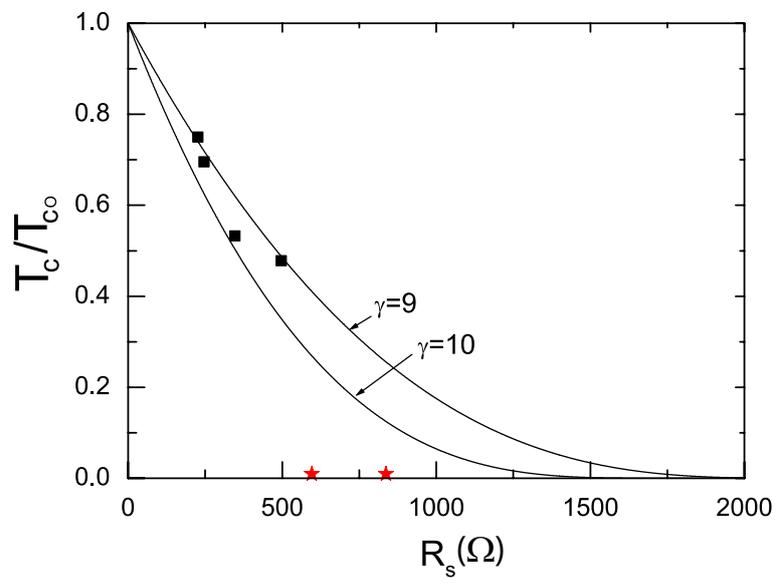

Delacour et al. Figure 7

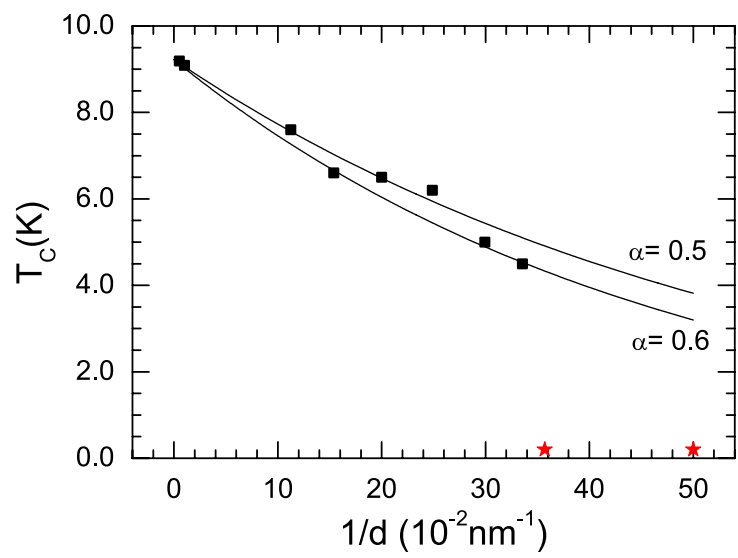

Delacour et al. Figure 8